\documentstyle[11pt]{article}
\begin{document}
\begin{flushright}SJSU/TP-98-16\\ January 1998\end{flushright}
\vspace{1.7in}
\begin{center}\Large{\bf Yet another comment on \\ ``Nonlocal
           character of quantum theory'' \\} 
\vspace{1cm}
\normalsize\ J. Finkelstein\footnote[1]{
        Participating Guest, Lawrence Berkeley National Laboratory\\
        \hspace*{\parindent}\hspace*{1em}
        e-mail: JLFinkelstein@lbl.gov}\\
        Department of Physics\\
        San Jos\'{e} State University\\San Jos\'{e}, CA 95192, U.S.A
\end{center}
\begin{abstract}
    There has been considerable discussion of the claim by Stapp \cite{S1}
    that quantum theory is incompatible with locality.  In this note I
    analyze the meaning of some of the statements used in this discussion.
\end{abstract}
\newpage
Stapp  has claimed to have proven that quantum theory is incompatible with
relativistic causality \cite{S1}.  This claim has been criticized by 
Unruh \cite{U} and by Mermin \cite{M1,M2}, 
and Stapp has replied to this criticism, in  \cite{S2,S3,S4}.

Stapp's proof involves a statement (to be explained below) which I will
denote by $S(L2)$.  Stapp claims to show that $S(L2)$ is true, that an
analogous statement $S(L1)$ is false, and that this difference in the
truth-values of $S(L2)$ and $S(L1)$ constitutes a violation of a 
locality condition he calls LOC2.  Unruh and Mermin state that they
agree that $S(L2)$ is true; thus the dispute between Stapp and his two critics
would seem to involve the relation between $S(L2)$ and $S(L1)$,
and, in particular, the meaning and applicability of LOC2. However, as I
will point out in this note, the agreement on $S(L2)$ is illusory; the
version of $S(L2)$ agreed to by Unruh and Mermin does not have the same
meaning as does $S(L2)$ as understood by Stapp. 
 In this note I will discuss the meaning of $S(L2)$.   

In ref. \cite{S1}, Stapp considers two particles in the (entangled) Hardy
state \cite{H}, on which measurements can be performed in two 
spacelike-separated regions called Right and Left.  
On the Right, measurement is
made of either of two non-compatible properties called $R1$ and $R2$; 
similarly, on the Left either $L1$ or $L2$ is measured. The result of any
given measurement is either $+$ or $-$.  The statement I call $S(L2)$ 
is the statement that, if $L2$ is measured, a statement called $S$ is true;
in symbols
\begin{equation}
S(L2):=[L2 \Rightarrow S]
\end{equation}
(similarly, $S(L1):=[L1 \Rightarrow S])$, where $S$ is defined by
\begin{eqnarray}
S &:= & \mbox{``If $R2$ is measured and yields the result $+$, then} 
\nonumber \\
& &\mbox{if $R1$ had been measured it would have yielded the result $-$''}
\end{eqnarray}

Now, what exactly does $S$ mean? In this (counterfactual) statement, one is
describing an ``actual'' world, in which $R2$ is measured, and a
``hypothetical'' world in  which, instead, $R1$ is measured; $S$ then
is the assertion that, in the hypothetical world, the result of $R1$ would
necessarily be $-$.  For this to make sense, it is necessary for the
hypothetical world to be specified more fully. {\em Roughly} speaking, the
idea is to specify that, except for the replacement of $R2$ by $R1$,
the hypothetical world agrees closely  with the actual world;
$S$ then is the assertion that, in {\em every} world that fits the 
specification, the result of $R1$ is $-$. And so to complete the 
definition of $S$, it is necessary to specify exactly in which ways the 
hypothetical world is required to agree with the actual world.

Here is one way to make the specification: to demand that the
hypothetical world agree with the actual world on all events which are
not in the invariant future of the measurement on the Right (that is,
on all events either spacelike-separated from, or else on or within the
backward lightcone from, that measurement). Let $F$ denote this set of
events (which are behind the {\bf F}orward lightcone),
and then define $S_F$ to be this version of $S$:
\begin{eqnarray}
  S_{F} & := & \mbox{``If $R2$ is measured and yields the result $+$, then}
                                         \nonumber \\
        &  & \mbox{in every world which agrees with the actual world on $F$}
                                         \nonumber \\
        &  & \mbox{(in particular, which agrees with the actual Left result)}
                                        \nonumber \\
        &  &\mbox {and in which $R1$ rather than $R2$ is measured,}
                                         \nonumber \\
        &  &\mbox{the result of $R1$ is $-$.''}
\end{eqnarray}
It is not important whether or not the definition (3) agrees with what
most people would mean by the statement $S$; what {\em is} important is
to have a clearly-defined definition for $S$.  Of course, (3)
might be motivated by physical principles such as causality, and those
principles might be relevant in proving that $S_F$ is true in some
particular case; nevertheless, (3) is just a definition, and we are free
to chose a different definition if we like. Here is an example of a different
definition: we could specify that the hypothetical world must agree with
the actual world on all events in the invariant past of the measurement
on the Right.  Let $B$ denote that set of events (behind the {\bf B}ackward
lightcone), and $S_B$ that version of $S$:
\begin{eqnarray}
  S_{B} & := & \mbox{``If $R2$ is measured and yields the result $+$, then}
                                         \nonumber \\
        &  &  \mbox{in every world which agrees with the actual world on $B$}
                                         \nonumber \\
        &  & \mbox {and in which $R1$ rather than $R2$ is measured,}
                                         \nonumber \\
        &    &\mbox{the result of $R1$ is $-$.''}
\end{eqnarray}  
$S_F$ and $S_B$ are not identical statements  (truth of $S_B$ implies truth
of $S_F$, but not the other way around).  No physical principle can tell
us whether either (3) or (4) give the ``correct'' meaning for $S$,
since definitions cannot be ``correct''.  I certainly do {\em not}
mean to assert that $S_B$ captures the usual meaning of $S$, nor do I 
mean to suggest $S_B$ as a useful alternative to $S_F$.  I have introduced
$S_B$ merely to emphasize that, to be unambiguous, statement $S$ must
include a specification of the ways in which the hypothetical world
is required to agree with the actual world, and to suggest that it is
a good idea to spell out that specification as completely as possible.  

So far, I have discussed the {\em meaning} of statement $S$; 
for that discussion,
the quantum state of the two particles was completely irrelevant. To discuss
the conditions under which $S$ is {\em true}, it will be necessary to remember 
that the particles are in the Hardy state \cite{H}. 
Let $S_{F}(L2)$ denote $S(L2)$ in which $S$ is understood
as $S_F$; that is,
\begin{equation}
S_{F}(L2) := [L2 \Rightarrow S_{F}],
\end{equation} 
with analogous definitions for $S_{B}(L2)$, $S_{F}(L1)$, and$S_{B}(L1)$.
It happens that, for particles in the Hardy state, 
$S_{F}(L2)$ is true, and $S_{B}(L2)$, $S_{F}(L1)$, and $S_{B}(L1)$
are all false.
\begin{itemize}
\item  To see that $S_{F}(L2)$ is true, note that a quantum 
calculation\footnote [2]
{The quantum predictions for the measurements we are discussing, for
particles in the Hardy state, are presented in refs. \cite{S1,U,M1}.}
shows, for the {\em actual} world of $S_{F}(L2)$ (in which $L2$ and
$R2$ are measured and the result of $R2$ is $+$), that the result of
$L2$ is necessarily $+$.  Since $S_F$ constrains the hypothetical
world to agree with the actual world on the Left, in the hypothetical
world that result is also $+$.  Then a quantum calculation for this
hypothetical world (in which $L2$ and $R1$ are measured, and the result
of $L2$ is $+$) requires the result of $R1$ to be $-$. Note that no locality
assumption is needed here; the truth of $S_{F}(L2)$ follows simply from its
definition and the quantum properties of the Hardy state.
\item  To see that $S_{B}(L2)$ is false, note that, again, in the
{\em actual} world the result of $L2$ is $+$.  Now, however, we are free to
consider a {\em hypothetical} world in which the result of $L2$ is $-$,
and in such a world ($L2$ and $R1$ measured, result of $L2$ is $-$),
a quantum calculation reveals a non-zero probability for the result of
$R1$ to be $+$.  Thus there is a hypothetical world consistent with the
specification of $S_{B}(L2)$ in which the result of $R1$ is $+$;
therefore $S_{B}(L2)$ is false.  Now it may seem strange to allow a 
hypothetical world in which the result of $L2$ is $-$, while in the
actual world that result is $+$; after all, how could the decision to
measure $R1$ rather than $R2$ change the result on the Left?  It may
help to remember that a hypothetical world is, {\it ipso facto},
not the same as the actual world.  If I choose to talk about a
hypothetical world, then I get to choose what world to talk about;
if I (perhaps foolishly) were to adopt $S_B$ as representing the
meaning of $S$, then I {\em could} find an allowed hypothetical world
in which the result of $R1$ is $+$.
\item For completeness, we can see that $S_{F}(L1)$ is false by noting
that in the actual world ($L1$ and $R2$ measured, result of $R2$ is $+$),
it is allowed that the result of $L1$ be $-$. Then in the hypothetical
world ($L1$ and  $R1$ measured, result of $L1$ is $-$), the result
of $R1$ is allowed to be $+$. Finally, since $S_{F}(L1)$ is false,
$S_{B}(L1)$ must be false also.
\end{itemize}
Again, I am certainly {\em not} advocating adopting $S_B$ to represent
the meaning of $S$.  I am suggesting that, whatever we wish $S$ to mean,
it is useful to spell out that meaning explicitly, by specifying the ways
in which the hypothetical world is required to agree with the actual world.

Mermin, in ref. \cite{M1}, denotes by (I) the statement 
here called $S(L2)$. From his discussion 
of why he considers this statement to be true, it is
clear that Mermin requires the hypothetical world to agree with the actual
world on all events which, in some frame, occur earlier than the measurement
on the Right.  Since for every event in the set we have called $F$ there
is a frame in which that event does precede the measurement on the Right, 
Mermin's understanding of $S(L2)$ coincides with $S_{F}(L2)$.  Although Unruh
\cite{U} does not give  general criteria for the interpretation of
counterfactuals, in his discussion of statement $S(L2)$ (which appears
in eqs. 12 and 13 of ref. \cite{U}) he strongly emphasizes that this
statement must be understood as requiring that the result of $L2$ be $+$,
and that is the aspect of $S_{F}(L2)$ which is relevant for all the 
further discussion of locality.  Stapp\cite{S1,S2,S3,S4}, however, 
evidently interprets $S(L2)$ differently. For him, the meaning of $S$
(as opposed to the conditions required for its proof) involves only
the Right region; he writes, for example \cite{S3}
``This `meaning' of statement $S$
is strictly in terms of a relationship between the possibilities for the 
outcomes of alternative possible experiments both of which are confined
to the region R(ight).'' So statement $S$ for Stapp is not the same
as $S_F$ (and incidentally, not the same as $S_B$ either); thus statement
$S$ is used to mean different things by Stapp and by his two critics.

Of course, the issue is not which is the ``correct'' meaning of statement
$S$---that is, after all, just a matter of definition.\footnote [3]
{Humpty Dumpty has remarked \cite{HD} ``When {\em I} use a word it means
just what I choose it to mean---neither more nor less.''}  The issue is
whether quantum mechanics is incompatible with relativistic locality.
But it is difficult to discuss {\em that} issue without 
unambiguous and agreed-upon definitions for the statements which are
being discussed.

\vspace{1cm}
Acknowledgement:
I have benefited from conversations with (but not necessarily agreement
from) Henry Stapp.
I would also like to acknowledge the hospitality of the
Lawrence Berkeley National Laboratory.

\end{document}